\documentclass[12pt]{article}

\makeatletter
\def\appendix{\par\clearpage
  \setcounter{section}{0}
  \setcounter{subsection}{0}
  \@addtoreset{equation}{section}
  \def\@sectname{Appendix~}
  \def\theequation{\thesection.\arabic{equation}}
  \def\thesection{\Alph{section}}}
\makeatother

\usepackage{amssymb}
\usepackage{graphicx}
\usepackage[sort&compress]{natbib}
\newcommand{\bea}{\begin{eqnarray}}
\newcommand{\eea}{\end{eqnarray}}

\begin{document}

\begin{titlepage}

\phantom{.}
\vspace{-3cm}

\hfill December 2005

\hfill ITP-2E-05

\vskip 0.3cm

\centerline{\bf Asymptotic neutrino-nucleon cross section and saturation effects}

\vskip 0.3cm

\centerline{R.~Fiore$^{a\dagger}$, L.L.~Jenkovszky$^{b\ddagger}$,
A.V.~Kotikov$^{c\star}$, F.~Paccanoni$^{d\ast}$,
A.~Papa$^{a\dagger}$}

\vskip 0.2cm

\centerline{$^{a}$ \sl  Dipartimento di Fisica, Universit\`a della Calabria}
\centerline{\sl Istituto Nazionale di Fisica Nucleare, Gruppo collegato di Cosenza}
\centerline{\sl I-87036 Arcavacata di Rende, Cosenza, Italy}

\centerline{$^{b}$ \sl  Bogolyubov Institute for Theoretical Physics}
\centerline{\sl Academy of Sciences of Ukraine}
\centerline{\sl UA-03143 Kiev, Ukraine}

\centerline{$^{c}$ \sl Bogolyubov Laboratory of Theoretical Physics}
\centerline{\sl Joint Institute for Nuclear Research}
\centerline{\sl RU-141980 Dubna, Russia}

\centerline{$^{d}$ \sl Dipartimento di Fisica, Universit\`a di Padova}
\centerline{\sl Istituto Nazionale di Fisica Nucleare, Sezione di Padova}
\centerline{\sl via F. Marzolo 8, I-35131 Padova, Italy}

\vskip 0.1cm

\begin{abstract}
In this paper we present a simple analytic expression for the (spin-averaged)
neutrino-nucleon cross section for ultra-high energies at twist-2, obtained as the
asymptotic limit of our previous findings. This expression gives values for
the cross section in remarkable numerical agreement with the previous numerical 
evaluation in the energy region relevant for forthcoming neutrino experiments.
Moreover, we discuss the role and the relevance of saturation and recombination effects 
in our approach, in comparison with other recent suggestions.
\end{abstract}

\vskip 0.1cm
\hrule

\vfill

$
\begin{array}{ll}
^{\dagger}\mbox{{\it e-mail address:}} &  \mbox{fiore,~papa@cs.infn.it} \\
^{\ddagger}\mbox{{\it e-mail address:}} & \mbox{jenk@bitp.kiev.ua} \\
^{\star}\mbox{{\it e-mail address:}} & \mbox{kotikov@thsun1.jinr.ru} \\
^{\ast}\mbox{{\it e-mail address:}} & \mbox{paccanoni@pd.infn.it} 
\end{array}
$

\end{titlepage}
\eject
\newpage

\section{Introduction}

Neutrino astronomy holds enormous scientific potential and
prospects for its development are much better at high energies
since the neutrino-nucleon cross section and the angular
resolution increase with energy~\cite{LM}. These features give the
opportunity to use large natural target media like ice and
atmosphere as detectors. The ultra-high energy neutrino-nucleon
cross section will become soon an important ingredient in the
interpretation of the results in these experiments. 

Above the energy at which the neutrino interaction length is
approximately equal to the diameter of the Earth, $\simeq 40$ TeV,
no experimental constraints exist on the neutrino cross section.
However, parton distribution functions fitted to the HERA data and
Standard Model constrain theoretical predictions~\cite{FKR,GAN,KMS,FAL} 
obtained with different models and all cross sections are remarkably consistent 
at the highest energies~\cite{MR}. In some models~\cite{KK,MVM,FAL2} the
introduction of non-linear screening effects produces only mild
changes and the aforesaid remarkable agreement between cross
sections survives; according to other models~\cite{JJM}, all-twist
formulation of QCD evolution equations~\cite{BAL,JJLL,HJ} entails a
drastic change on the cross sections in the region where unitarity
effects become important. Geometric scaling~\cite{SGK}, that is a
consequence of the all-twist formulation of QCD, and a precise
form of the dipole cross section~\cite{MTT} in the geometric
scaling region are assumed in the approach of Ref.~\cite{JJM}. We will try
to explain roughly these concepts since they will be important in
the following. 

In the color dipole picture~\cite{AHM} the virtual photon, or the gauge 
bosons $W$ and $Z$ for neutrino scattering, creates a $q \bar{q}$ pair, 
or a ``color dipole". At high energies, or at small $x$, the exchange of 
gluons between the nucleon and the color dipole becomes more and more 
important and the gluon density in the nucleon increases with energy. The
quark-antiquark dipole has zero color charge and the interaction
with the gluons in the nucleon will depend on the dipole size. If
the size is very small the dipole will not interact with gluons,
but there will be a length $R_0(x)=1/Q_s(x)$ such that, for dipole
sizes greater than $R_0(x)$, the scattering cross section will be
perceptible. $R_0(x)$ is called the saturation radius and $Q_s(x)$
the saturation scale. Geometrical scaling refers to the dependence
of the dipole cross section $\hat{\sigma}(x,r)$, where $r$ is the
transverse separation of the quarks in the $q\bar{q}$ pair, from
only one dimensionless variable $r/R_0(x)$. As a consequence the
$\gamma^*\,p$ cross section, for example, becomes a function of
one dimensionless variable $\tau=Q^2R_0^2(x)$~\cite{SGK}. At small
$x$, and hence at high gluon density in the nucleon, $R_0(x)$
is small and the saturation scale is large. 

We return now to the neutrino-nucleon cross section. According to
Ref. \cite{JJM}, at neutrino energies $E_{\nu}>O\left(10^{12}\right)$ GeV 
geometric scaling holds and neutrino-nucleon cross sections are enhanced 
by a large factor. A higher cross section in this energy region could have 
important consequences for neutrino astronomy~\cite{KW}. Hence, a comparison 
with approaches, where geometric scaling and saturation scale have a
different validity range and interpretation, becomes interesting.

In a series of papers~\cite{ZZR,ZAL} modified evolution equations were 
suggested that include twist-4 gluon recombination corrections. These equations 
are quoted as ``modified DGLAP equations" because of their similarity with the 
original evolution equations~\cite{DGL,AP}: gluon recombination is evaluated 
in Refs.~\cite{ZZR,ZAL} with the same technique adopted in Ref.~\cite{AP} 
at twist-2. The physical picture of the deep inelastic scattering (DIS) process 
is different from the color dipole one, since now the virtual gauge boson 
explores the parton distribution in the nucleon. In the leading logarithmic 
approximation, and at twist-2 level, the two pictures give the same results but, 
at higher twist, this equivalence is lost~\cite{SZ} since approximations are
different in different frames. According to Ref.~\cite{SZ}, the most
important reason for this difference is that the color dipole
approach extracts the splitting probabilities incoherently and
neglects the coherence among different subpartonic amplitudes.

It has been shown in Ref.~\cite{AKUW} that the Balitsky and Kovchegov non-linear
evolution equation~\cite{BAL,YUK} leads to saturation of the scattering amplitude,
but does not necessarily unitarize the total cross section. The violation of 
unitarity depends on the nature of the target and the reasons for its
appearance can be seen both in the target rest frame or from the
point of view of the evolution of target fields. Interactions
between the dipoles in the projectile wave-function are neglected
in the color dipole approach and, on the other side, the evolution
in the target is driven from incoherent color sources. Unitarity
is violated since the dipole interacts with the long range Coulomb
field created by a large number of incoherent color sources in the
target. The proof in Ref.~\cite{AKUW} holds for a strong interacting
particle colliding on a hadronic target, but for DIS the situation changes 
for the worse since the DIS cross section acquires an extra power of the rapidity. 
As noticed in Ref.~\cite{AKUW}, the condition that the total color charge must be
zero, in a region of finite size (e.g. the proton), introduces
correlations, and coherence, among the sources of the color
charges and can lead to a unitary evolution. These considerations
justify the interest for a comparison between the two different
approaches to saturation and its consequences for neutrino
astronomy. 

In our previous paper~\cite{FAL2} we have obtained a simplified solution of 
the non-linear evolution equations of Refs.~\cite{ZZR,ZAL} at small $x$. 
In this paper we will first show that it is possible to simplify further the
integrals leading to the charged current neutrino-nucleon cross
section (Section~2). The answer at twist-2 will be analytical and take a very
simple form. In this simplified approach it becomes easy to
prove the well known statement that, at asymptotic energies, only
the values of $Q^2\sim M_W^2$ contribute to the cross section~\cite{ABS,KMS}
and an estimate of the neglected terms will be given (Section~3).

With the parameters, in the input parton distributions, fixed at the values obtained
in Ref.~\cite{FAL2}, we will then discuss the saturation phenomenon in
the approach of Refs.~\cite{ZZR,ZAL}. Similarities and differences
with the approach in Ref.~\cite{JJM} will be finally emphasized (Section~4).

\section{Asymptotic form of the neutrino-nu\-cleon cross section}

The starting point is the inclusive, spin-averaged cross section
for the neutrino interaction with an isoscalar nucleon target,
$N=(\mbox{neutron}+\mbox{proton})/2$, in the process
\begin{equation}
\nu_{\mu}(k)+N(p) \rightarrow \mu(k')+X(p')\;,  \label{n0}
\end{equation}
where parentheses enclose the four-momenta of the particles
participating to the scattering. The transformation of the 
neutrino to a charged muon labels the event (\ref{n0}) as a
``charged current" event and the charged current cross section can
be expressed in terms of the nucleon structure function as
\begin{eqnarray}
\left(\frac{d\sigma}{dx\,dy}\right)^{\nu} & = &
\frac{G_F^2ME_{\nu}}{\pi} \left(
\frac{M_W^2}{Q^2+M_W^2}\right)^2 \nonumber \\
&\times&
\left\{\left(1-y-\frac{Mxy}{2E}\right)F_2^{\nu}+\frac{y^2}{2}\,2xF_1^{\nu}+
\left(1-\frac{y}{2}\right)yxF_3^{\nu}\right\}. \label{n1}
\end{eqnarray}
For anti-neutrino charged-current processes one must change the
sign in front of $F_3^{\nu}$, while the changes necessary in order to
describe neutral-current neutrino interactions can be found in the
literature~\cite{Buras}. In Eq.~(\ref{n1})
$F_i^{\nu}=F_i^{\nu}(x,Q^2)$ for $i=1,\,2,\,3$, $G_F$ is the Fermi
constant, $M$ is the nucleon mass and $M_W$ is the $W$-boson mass.
The scaling variables $x$ and $y$ are defined as
\begin{equation}
x=-\frac{q^2}{2p\cdot q},\;\;\;\;\;y=\frac{p\cdot q}{p\cdot k}
\label{n2}
\end{equation}
and $Q^2=-q^2$. 

At high energies the relation between
the variable $y$ and the Bjorken variable $x$ can be approximated as
\begin{equation}
y=\frac{Q^2}{x(s-M^2)}\simeq \frac{Q^2}{xs}\;, \label{n3}
\end{equation}
where $s=(k+p)^2$ is the square of the c.m. energy for the
neutrino-nucleon  scattering. The laboratory neutrino energy
$E_{\nu}=(s-M^2)/(2M)$ is approximately equal to $s/(2M)$ in the region we
consider. 

The main purpose of this paper is to evaluate the asymptotic behavior of the 
total cross section for the charged-current neutrino-nucleon process
\begin{equation}
\sigma^{\nu N\,(CC)}=\int_0^1
dx\,\int_0^1dy\,\left(\frac{d\sigma}{dx\,dy}\right)^{\nu}\;.
\label{n4}
\end{equation}
In the following, we will limit ourselves to the leading order
corrections to the simple parton model and hence all parton model
formulas remain unchanged except that the parton distributions depend now
on $x$ and $Q^2$ and not only on $x$. In particular, the
Callan-Gross relation, $F_2=2xF_1$ or $F_L=0$, holds in leading
order. By imposing a lower cut in $Q^2$, $Q^2=Q_0^2$, we rewrite
Eq.~(\ref{n4}) as
\begin{displaymath}
\sigma^{\nu
N\,(CC)}\simeq\frac{1}{2ME_{\nu}}\int_{Q_0^2}^s\,dQ^2\,
\int_{Q^2/s}^{1}\,\frac{dx}{x}\left(\frac{d\sigma}{dx\,dy}\right)^{\nu}\;,
\end{displaymath}
where $y=Q^2/(xs)$ in the differential cross section, according to Eq.~(\ref{n3}). 
Then, the $F_2$ contribution to the total cross section can be written in the form
\begin{displaymath}
\bar{\sigma}^{\nu N}\equiv\frac{\sigma^{\nu
N}+\sigma^{\bar{\nu}N}}{2}
\end{displaymath}
\begin{equation}
=\frac{G_F^2}{2\pi}\int_{Q_0^2}^s\,dQ^2\left(
\frac{M_W^2}{Q^2+M_W^2}\right)^2
\int_{Q^2/s}^{1}\,\frac{dx}{x}\;\frac{1+(1-Q^2/(xs))^2}{2}F_2^{\nu}(x,Q^2)\;,
\label{n5}
\end{equation}
where we have neglected the term $-Mxy/(2E)$ in Eq.~(\ref{n1}). Since we are mainly 
interested in the asymptotic $s$ behavior, some simplifications are possible and,
when $s$ is much larger than all the scales appearing in Eq.~(\ref{n5}), in particular 
$s\gg M_W^2$, 
\begin{enumerate}
\item the contribution of $xF_3$ can be neglected and $\bar{\sigma}^{\nu
N}\simeq \sigma^{\nu N} \simeq \sigma^{\bar{\nu}N}$;
\item the inequality $s\gg Q^2$ holds because the factor $(M_W^2/(Q^2+
M_W^2))^2$ limits the  $Q^2$ integration region: as will be shown later, 
the upper limit of the $Q^2$ integral becomes proportional to $M_W^2$.
\end{enumerate}

As in Ref.~\cite{FAL2}, we write the isoscalar structure function in terms 
of the parton distribution functions and, at leading order, we get
\begin{displaymath}
F_2^{\nu}(x)\simeq x\,u(x)+x\,\bar{u}(x)+x\,d(x)+x\,\bar{d}(x)+
2x\,s(x)+2x\,c(x)+\ldots \;,
\end{displaymath}
where the dots stand for the $b$- and $t$-quark PDFs and we have assumed
$s(x)=\bar{s}(x)$ and $c(x)=\bar{c}(x)$. We denote, in the following, by
$f_q(x,Q^2)$ the sea quark distribution $xS(x,Q^2)$ at twist-2 and
by $f_g(x,Q^2)$ the gluon distribution $xG(x,Q^2)$ in the same approximation
and use the notation $f_q^{full}(x,Q^2)$ for the sea quark distribution
modified by the introduction of gluon recombination at twist-4~\cite{ZAL}.

Setting $u=Q^2/s$, we can write the last integral in Eq.~(\ref{n5})
in the form
\begin{equation}
\int_u^1 \frac{dx}{x} \left[1-\frac{u}{x}+\frac{u^2}{2x^2}
\right]\,f_q^{full}(x,Q^2)\;, \label{n6}
\end{equation}
which is a Mellin convolution that, in moment space, becomes
\begin{eqnarray}
&& \int_u^1\,\frac{dx}{x}\left(1-\frac{u}{x}+\frac{u^2}{2x^2}
\right)\,f_q^{full}(x,Q^2) \nonumber \\ && ~~\stackrel{M}{\to} ~~
\left(\frac{1}{n-1}-\frac{3}{4}+O(n-1)\right)\,f_q^{full}(n,Q^2)\;,
\label{n7}
\end{eqnarray}
since, as noticed before, large $Q^2$ contributions are strongly
suppressed by the factor $(1+Q^2/M_W^2)^{-2}$ and $u$ can be
considered small. The result in Eq.~(\ref{n7}) can be obtained by
considering the relation
\begin{displaymath}
\int_0^1u^{n-2}du\int_u^1\frac{dx}{x}M_1\left(\frac{u}{x}\right)
M_2(x)=M_1(n)M_2(n)\;,
\end{displaymath}
where
\begin{eqnarray*}
M_1(n) &=& \int_0^1dt\,t^{n-2}\left(1-t+\frac{t^2}{2}\right) \\
&=& \frac{1}{n-1}-\frac{1}{n}+\frac{1}{2(n+1)}\sim \frac{1}{n-1}
-\frac{3}{4}+O(n-1)
\end{eqnarray*}
and
\begin{displaymath}
M_2(n)=\int_0^1dx\,x^{n-2}f_q^{full}(x,Q^2)\equiv
f_q^{full}(n,Q^2).
\end{displaymath}
With the definition
\begin{equation}
\left(\frac{1}{n-1}-\frac{3}{4}\right)f_q^{full}(n,Q^2)
~~\stackrel{M^{-1}}{\to} ~~ g(u,Q^2)\;, \label{n8}
\end{equation}
Eq.~(\ref{n5}) becomes
\begin{equation}
\bar{\sigma}^{\nu N}=\frac{G_F^2}{2\pi}
\int_{Q_0^2}^s\,dQ^2\left(1+\frac{Q^2}{M_W^2}\right)^{-2}
g(Q^2/s,Q^2). \label{n9}
\end{equation}

\section{Twist-2 contributions to the cross section}

To begin with we consider our approximate twist-2 solution for the
structure function $F_2(x,Q^2)$, that is the first term $f_q$ in
\begin{displaymath}
f_q^{full}=f_q+T_q\;,
\end{displaymath}
where $T_q$ represents the twist-4 gluon recombination
corrections. It is important to ensure the accuracy of the
approximation (\ref{n9}) in the simplest case and to explore the
possibility of further simplifications. From our previous work
\cite{FAL,FAL2}, where the method introduced in Refs.~\cite{LY,M79,AVK,AKP} and
used also in Ref.~\cite{IKP} was adopted, we have
\begin{eqnarray}
f_q(n,Q^2) &=& \frac{1}{n-1}A_q^-e^{-d_-(1)\,t} \nonumber \\
&+& A_q^+\,\sum_{k=0}^{\infty}\frac{1}{(k+1)!}\left(\frac{ -
\hat{d}_{gg}t}{n-1}\right)^{k+1}e^{-\bar{d}_+(1)\,t}\;, \label{n10}
\end{eqnarray}
where
\begin{displaymath}
t=\ln\left[\frac{\alpha_s(Q_0^2)}{\alpha_s(Q^2)}\right],
\end{displaymath}
$\hat{d}_{gg}=-12/\beta_0,\;\bar{d}_-(1)=16f/(27\beta_0)$ and
$\bar{d}_+(1)=1+20 f/(27\beta_0)$, with $\beta_0=11-2 f/3$ and $f$
the number of flavors. Introducing the new variables
\begin{equation}
\sigma_u = 2\sqrt{-\hat d_{gg} t \ln(1/u)},~~~~~\rho_u =
\frac{\sigma_u}{2\ln(1/u)}, \label{n11}
\end{equation}
we find the following expression for the twist-2 contribution to $g(u,Q^2)$:
\begin{eqnarray}
g^{(2)}(u,Q^2) &\simeq&
A_q^-\left[\ln\frac{1}{u}-\frac{3}{4}\right]
e^{-d_-(1) t} + \nonumber \\
&+& A_q^+\left[I_0(\sigma_u)-\frac{3}{4}\rho_uI_1(\sigma_u)\right]
e^{-\bar{d}_+(1)t} \;.\label{n12}
\end{eqnarray}

The simplest contribution to Eq.~(\ref{n9}) coming from $g^{(2)}(u,Q^2)$ is
$$-\frac{3}{4}A_q^-\,e^{-d_-(1)t}\;,$$
where we can put $\exp[-d_-(1)t]=[\ln(Q_0^2/\Lambda^2)/
\ln(Q^2/\Lambda^2)]^{d_-(1)}$. A basic integral appearing in Eq.~(\ref{n9}) is 
then the following:
\begin{equation}
{\cal I}_1(d)=\int_{Q_0^2}^s\,dQ^2 \left(1+\frac{Q^2}{M_W^2}
\right)^{-2}\frac{1}{[\ln(Q^2/\Lambda^2)]^d}\;. \label{n13}
\end{equation}
The proof that, when $s\to \infty$,
\begin{equation}
{\cal I}_1(d) \rightarrow M_W^2\left[\ln\left(
\frac{M_W^2}{\Lambda^2}\right)\right]^{-d}, \label{n14}
\end{equation}
neglecting terms proportional to $[\ln(M_W^2/\Lambda^2)]^{-2-d}$,
is rather long, but important and will be presented in Appendix~A.

It is not difficult to generalize the proof of Eq.~(\ref{n14}) to an 
expression of the form
\begin{displaymath}
{\cal I}_j=\int_{Q_0^2}^s\,dQ^2 \left(1+\frac{Q^2}{M_W^2}
\right)^{-2}\;j[\ln(Q^2/\Lambda^2)]\;, \label{n132}
\end{displaymath}
where $j[z]$ is a function that can be expanded in powers of
$1/z$, and prove that, at twist-2,
\begin{eqnarray}
\bar{\sigma}^{\nu N\;(2)} &\simeq & \frac{G_F^2}{2\pi}M_W^2
\left[A_q^-\left(\ln\frac{s}{M_W^2}\;-\frac{3}{4}\right)
e^{-d_-(1)\hat{t}} \right. \nonumber \\ &+& \left.
A_q^+\left(I_0(\hat{\sigma})
-\frac{3}{4}\hat{\rho}I_1(\hat{\sigma})
\right)e^{-\bar{d}_+(1)\hat{t}} \right]\;, \label{n22}
\end{eqnarray}
where
\begin{displaymath}
\hat{t}=\ln\left[\frac{\alpha_s(Q_0^2)}{\alpha_s(M_W^2)}\right],
~~~~~\hat{\sigma} = 2\sqrt{-\hat d_{gg} \hat{t}
\ln(s/M_W^2)},~~~~~\hat{\rho} = \frac{\hat{\sigma}}{2
\ln(s/M_W^2)}. \label{n23}
\end{displaymath}
We notice that the expression for the twist-2 cross section is explicit. In Table~1 
and Fig.~1 this approximation is compared with the numerical determination obtained 
in Ref.~\cite{FAL2}
in the case of absence of recombination (see Fig.~4 of that paper). The values used
for the parameters $A_q$ and $A_g$ are 1.040(36) and 0.548(28), 
respectively~\footnote{We remind that $A_q^+=f(A_g+4 A_q/9)/9$ and 
$A_q^-= A_q$, for flat initial conditions~\cite{FAL2}.}. The approximate cross 
section given in Eq.~(\ref{n22}) nicely matches the numerical determinations of 
our previous work~\cite{FAL2} and those of Refs.~\cite{FKR,GAN,KMS}.

When the gluon recombination term is present, any attempt to simplify the 
problem becomes much more intricate but, as we will see in the next
Section, the discussion of the saturation limit can be done on the
basis of a simpler approach.

\begin{table}[ht]
\begin{center}
\begin{tabular}{||c|r|r||} 
\hline\hline 
$s$ [GeV$^2$] & ${\bar \sigma}_{\nu N}$ [cm$^2$], Ref.~\cite{FAL2} 
              & ${\bar \sigma}_{\nu N}$ [cm$^2$], Eq.~(\ref{n22}) \\ 
\hline\hline 
  $10^5$      & $9.75(39) \times 10^{-35}$ & $1.027(38) \times 10^{-34}$ \\ \hline 
  $10^6$      & $4.13(17) \times 10^{-34}$ & $3.65(15)  \times 10^{-34}$ \\ \hline  
  $10^7$      & $1.336(57)\times 10^{-33}$ & $1.112(46) \times 10^{-33}$ \\ \hline  
  $10^8$      & $3.75(16) \times 10^{-33}$ & $3.20(14)  \times 10^{-33}$ \\ \hline  
  $10^9$      & $9.67(42) \times 10^{-33}$ & $8.61(37)  \times 10^{-33}$ \\ \hline  
  $10^{10}$   & $2.33(10) \times 10^{-32}$ & $2.164(94) \times 10^{-32}$ \\ \hline  
  $10^{11}$   & $5.34(23) \times 10^{-32}$ & $5.11(22)  \times 10^{-32}$ \\ \hline  
  $10^{12}$   & $1.167(51)\times 10^{-31}$ & $1.143(50) \times 10^{-31}$ \\ \hline  
  $10^{13}$   & $2.45(11) \times 10^{-31}$ & $2.44(11)  \times 10^{-31}$ \\ \hline  
  $10^{14}$   & $4.97(22) \times 10^{-31}$ & $5.02(22)  \times 10^{-31}$ \\ \hline  
  $10^{15}$   & $9.80(43) \times 10^{-31}$ & $9.96(43)  \times 10^{-31}$ \\  
\hline\hline
\end{tabular}
\end{center}
\caption{Comparison between the twist-2 contribution to the neutrino-nucleon cross
section according to the numerical results obtained in Ref.~\cite{FAL2} (2nd column) 
and the asymptotic approximation given in Eq.~(\ref{n22}) (3rd column). 
The errors come, in both cases, from the uncertainties of the $A_q^{\pm}$ parameters.}
\label{tab1}
\end{table}

\begin{figure}[tb]
\centering
\includegraphics[width=\textwidth]{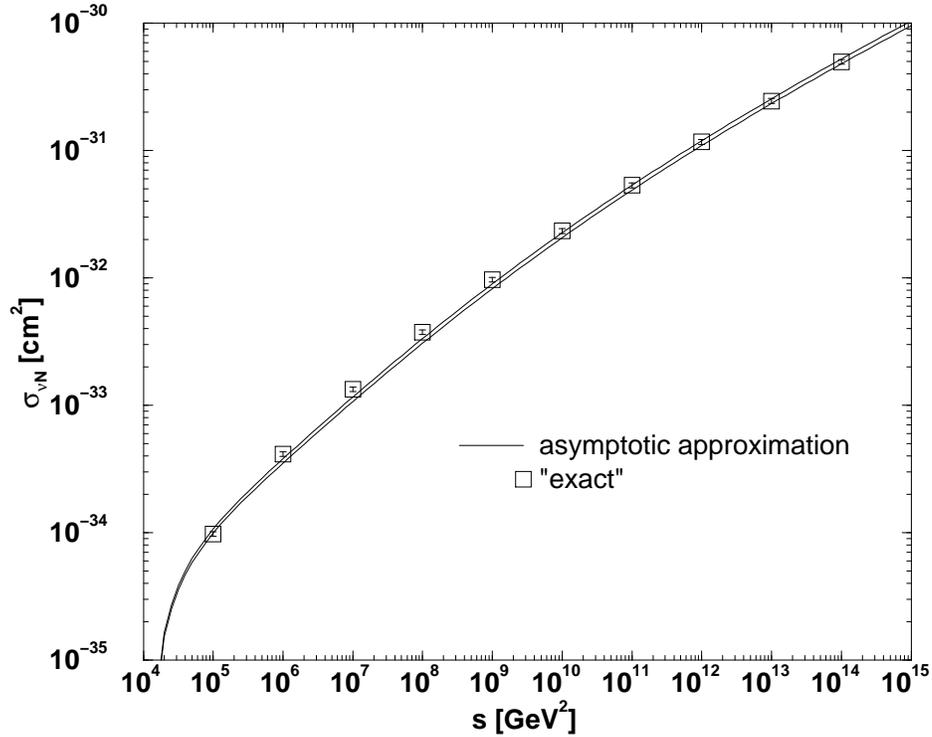}
\caption[]{\small Comparison between the twist-2 contribution to the neutrino-nucleon 
cross section according to numerical result shown in Fig.~4 of Ref.~\cite{FAL2} 
(squares) and the asymptotic approximation given in Eq.~(\ref{n22}) (continuous 
lines, representing the upper and the lower bounds at 1$\sigma$ level). 
The uncertainty comes, in both cases, from that of the $A_q^{\pm}$ parameters.}
\label{fig1}
\end{figure}

\section{Saturation and recombination scales}

The saturation scale $Q_S^2(x)$ indicates the saturation limit and
is usually defined as~\cite{GLR,KMSR}
\begin{displaymath}
\left. \frac{d\;xG(x,Q^2)}{d\;\ln Q^2}\right|_{Q_S^2}=0\;\;\mbox{
and}\;\;\;\;\;\left. \frac{d\;xS(x,Q^2)}{d\;\ln
Q^2}\right|_{Q_S^2}=0
\end{displaymath}
or, equivalently~\cite{ZAL},
\begin{displaymath}
W_S\equiv \left. \frac{\mbox{non-linear terms}}{\mbox{linear
terms}}\right|_{Q_S^2}=1 \;,
\end{displaymath}
which means that the non-linear recombination effect in the MD-DGLAP equation 
fully balances the linear splitting effect.
The recombination scale $Q_R^2$, introduced in Ref.~\cite{ZAL}, is defined
as
\begin{displaymath}
W_R\equiv \left. \frac{\mbox{non-linear terms}}{\mbox{linear
terms}}\right|_{Q_R^2}=\alpha_s[Q_R^2(x)]\;,
\end{displaymath}
which means that, near this scale, the higher-order recombination contributions 
cannot be neglected and should be included in the evolution, thus making the 
evolution of the parton distributions from $Q_R^2$ to $Q_S^2$ much more complicated.

Since, according to Ref.~\cite{ZAL}, saturation and recombination appear at very 
low $x$ we are justified to use approximate relations like
\begin{equation}
\frac{dF_2^{\nu N}(x,Q^2)}{d\ln
Q^2}=\frac{df_q^{full}(x,Q^2)}{d\ln Q^2} \sim
\frac{df_q(x,Q^2)}{d\ln
Q^2}+\frac{\alpha_s^2}{Q^2}K\left(-\frac{17}{32}
f_g^2(x,Q^2)\right) \label{n23b}
\end{equation}
(see, for example, Eq.~(64) in Ref.~\cite{FAL2} and Appendix~B).
The saturation scale will be consequently defined from the equation
\begin{equation}
\left. \frac{\frac{df_q(x,Q^2)}{d\ln Q^2}}{\frac{17}{32} \frac{K}{Q^2}\,f_g^2(x,Q^2)}
\right|_{Q_S^2}=\alpha_s^2 \;, \label{n24}
\end{equation}
while the recombination scale satisfies
\begin{equation}
\left. \frac{\frac{df_q(x,Q^2)}{d\ln Q^2}}{\frac{17}{32} \frac{K}{Q^2}\,f_g^2(x,Q^2)}
\right|_{Q_R^2}=\alpha_s \;. \label{n25}
\end{equation}
We can use in these equations the values of the parameters obtained in the fit 
of ZEUS PDF~\cite{zeus} in order to have an idea of the behavior of the scale 
$Q_S^2$, and more importantly of the scale $Q_R^2$, with $x$. This is an interesting 
point since twist-4 recombination formulas hold near the recombination scale but, if 
we approach the saturation scale, higher order recombination contributions become 
significant~\cite{ZAL}. In other words, we identify the region where our formulas 
can be trusted. From Eqs.~(\ref{n24}) and (\ref{n25}), we can build a numerical 
table (see Table~2 and Fig.~2), using the results of our previous work~\cite{FAL2}, 
where, in particular, the $K$ parameter was set to 0.013.

\begin{table}[ht]
\begin{center}
\begin{tabular}{||c|c|c||} 
\hline\hline 
$Q^2$ [GeV$^2]$ & $x_{rec}$             & $x_{sat}$             \\ 
\hline\hline 
       5        & $1.35\times 10^{-8}$  & $3.27\times 10^{-11}$ \\ \hline 
      10        & $8.43\times 10^{-9}$  & $1.79\times 10^{-11}$ \\ \hline 
      50        & $7.49\times 10^{-10}$ & $9.14\times 10^{-13}$ \\ \hline 
     100        & $1.96\times 10^{-10}$ & $1.82\times 10^{-13}$ \\ \hline 
     200        & $4.59\times 10^{-11}$ & $3.20\times 10^{-14}$ \\ \hline
     500        & $5.87\times 10^{-12}$ & $2.78\times 10^{-15}$ \\ \hline 
    1000        & $1.14\times 10^{-12}$ & $4.00\times 10^{-16}$ \\ \hline 
    2000        & $2.06\times 10^{-13}$ & $5.39\times 10^{-17}$ \\ \hline 
    3000        & $7.37\times 10^{-14}$ & $1.62\times 10^{-17}$ \\ \hline 
    4000        & $3.52\times 10^{-14}$ & $6.82\times 10^{-18}$ \\ \hline 
    5000        & $1.97\times 10^{-14}$ & $3.46\times 10^{-18}$ \\ \hline 
   $M_W^2$      & $9.98\times 10^{-15}$ & $1.57\times 10^{-18}$ \\ 
\hline\hline
\end{tabular}
\end{center}
\caption{Recombination and saturation scales, defined according to Eqs.~(\ref{n24}) 
and (\ref{n25}), determined using the PDFs of our previous work~\cite{FAL2}.}
\label{tab2}
\end{table}

\begin{figure}[tb]
\centering
\includegraphics[width=\textwidth]{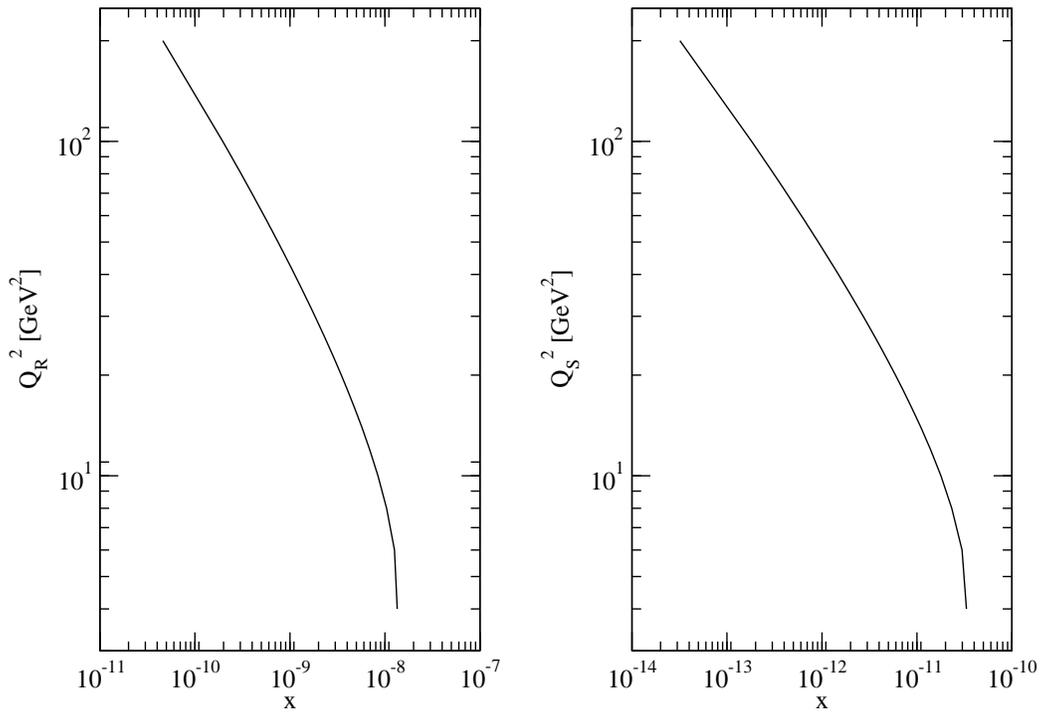}
\caption[]{\small Recombination (left) and saturation (right) scales, defined according to 
Eqs.~(\ref{n24}) and (\ref{n25}), determined using the PDFs of our previous 
work~\cite{FAL2}.}
\label{fig2}
\end{figure}

From this table we realize that
\begin{enumerate}
\item the function $Q_R^2(x)$ is quite similar to the one obtained in Ref.~\cite{ZAL}
with different values of the parameters (in particular the value of $K$ is quite 
different in our approach);
\item the recombination scale agrees with our findings for the $Q^2$ slope of 
$F_2^{\nu N}(x,Q^2)$;
\item the evaluation we did of the neutrino-nucleon cross section
is safe since the small-$x$ limit considered is well above the $x$-value 
associated with the recombination scale at $Q^2\sim M_W^2$; in other words 
higher order twists, besides twist-4, are not important.
\end{enumerate}

This last point gives sense to a comparison of our results with those
of Ref.~\cite{JJM}, where all twists were resummed, in the region of energy
we considered.

As far as anti-shadowing effects are concerned, they are important at values of $Q^2$
much smaller that $M_W^2$~\cite{ZAL}, therefore they are completely negligible in our
analysis.

In conclusion we find that our approach is sound and reliable. It has many 
points in common with the analysis of Ref.~\cite{JJM}: saturation presents itself
at very small $x$ and is not detectable, antishadowing is present in both approaches, 
but at different values of $Q^2$. Enhancement of ultra-high energy neutrino-nucleon 
cross section is not required in our calculation and this is due to the correlation 
among color sources in the target. Such correlation, and coherence, is absent
in the color dipole picture.

\vspace{0.5cm} {\bf \large Acknowledgments} L.J. and A.K. thank
the Departments of Physics of the Universities of Calabria and
Padova, together with the INFN Gruppo collegato di Cosenza and
Sezione di Padova for their warm hospitality and support. This work was
partially supported by the Ministero Italiano dell'I\-stru\-zio\-ne, dell'Universit\`a
e della Ricerca.

\appendix

\section{Appendix: Proof of Eq.~(\ref{n14})}

With the change of variable $t=\ln(Q^2/\Lambda^2)$ we can rewrite the 
integral~(\ref{n13}) in the following forms
\begin{eqnarray*}
{\cal I}_1(d) \!\!\!&=&\!\!\! \frac{M_W^4}{\Lambda^2}\!\int_{\ln(Q_0^2/
\Lambda^2)}^{\ln(s/\Lambda^2)} \frac{t^{-d}e^t\;dt}{\left(e^t+
M_W^2/\Lambda^2\right)^2} \\ 
\!\!\!&=&\!\!\! \frac{M_W^4}{\Lambda^2}\!\left[\int_{\ln(Q_0^2/
\Lambda^2)}^{\ln(s/\Lambda^2)} \frac{t^{-d}\;dt}{e^t+
M_W^2/\Lambda^2}-\frac{M_W^2}{\Lambda^2}\int_{\ln(Q_0^2/
\Lambda^2)}^{\ln(s/\Lambda^2)} \frac{t^{-d}\;dt}{\left(e^t+
M_W^2/\Lambda^2\right)^2}\right] \\ 
\!\!\!&=&\!\!\! \frac{M_W^4}{\Lambda^2}\!\left[\int_{\ln(Q_0^2/
\Lambda^2)}^{\ln(s/\Lambda^2)} \frac{t^{-d}e^{-t}\;dt}{1+
M_W^2e^{-t}/\Lambda^2}-\frac{M_W^2}{\Lambda^2}\int_{\ln(Q_0^2/
\Lambda^2)}^{\ln(s/\Lambda^2)} \frac{t^{-d}e^{-2t}\;dt}{\left(1+
M_W^2e^{-t}/\Lambda^2\right)^2}\!\right]\!.
\end{eqnarray*}
Setting $z=M_W^2/\Lambda^2$, we have
\begin{eqnarray}
{\cal I}_1(d) &=&
M_W^2z\left(1+z\frac{d}{dz}\right)\int_{\ln(Q_0^2/
\Lambda^2)}^{\ln(s/\Lambda^2)} \frac{t^{-d}e^{-t}\;dt}{1+z e^{-t}}
\nonumber \\ &\equiv & M_W^2z\left(1+z\frac{d}{dz}\right) {\cal
L}(d,z)\;. \label{n15}
\end{eqnarray}
${\cal L}(d,z)$ differs from the integral
\begin{displaymath}
\int_{\ln(Q_0^2/ \Lambda^2)}^{\infty} \frac{t^{-d}e^{-t}\;dt}{1+z
e^{-t}}
\end{displaymath}
by terms vanishing faster than $1/s$ in the asymptotic region
for the variable $s$. This result can be easily obtained by
expanding in series the integral in Eq.~(\ref{n15}) with respect to
its upper limit. Moreover, since $z$ is a large number ($\ln z\sim
12$ if $\Lambda=0.19$ GeV) another approximation becomes possible
and, neglecting terms proportional to $[\ln(z)]^{-2}$ with respect
to a constant term, we have
\begin{equation}
{\cal L}(d,z)\simeq \int_0^{\infty} \frac{t^{-d}e^{-t}\;dt}{1+z
e^{-t}}\;. \label{n16}
\end{equation}
The integral (\ref{n16}) can be expressed as an infinite sum
\begin{eqnarray}
{\cal L}(d,z) &\simeq & \sum_{n=0}^{\infty}
(-1)^nz^n\int_0^{\infty} dt\,t^{-d}e^{-(n+1)t} \nonumber \\ & = &
-\frac{1}{z}\Gamma(1-d)\,\sum_{n=1}^{\infty}\frac{(-z)^n}{n^{1-d}}
\label{n17}
\end{eqnarray}
and finally
\begin{equation}
{\cal L}\simeq -\frac{1}{z} \Gamma(1-d) F(-z,1-d)\;, \label{n18}
\end{equation}
where $F(z,2)$ is the Euler's dilogarithm. The analytical continuation of the series 
\begin{displaymath}
F(-z,1-d)=\sum_{n=1}^{\infty}\;\frac{(-z)^n}{n^{1-d}}
\end{displaymath}
is given by the Joncqui\`ere's relation~\cite{BAT} that, in our case, becomes
\begin{displaymath}
F(-z,1-d)=-e^{i(1-d)\pi}F(-1/z,1-d)+\frac{(2\pi)^{1-d}}{\Gamma(1-d)}
e^{i\pi (1-d)/2}\zeta\left(d,\frac{\ln(-z)}{2\pi i}\right).
\end{displaymath}
The variable $z=M_W^2/\Lambda^2$ is very large and an asymptotic
expansion for ${\cal L}$ follows from the asymptotic expansion of
the generalized Zeta function for $z\to \infty$,
\begin{eqnarray}
\zeta\left(d,\frac{\ln(-z)}{2\pi i}\right) &\rightarrow &
\frac{1}{\Gamma(d)}\left[\Gamma(d-1)\left(\frac{\ln(-z)}{2\pi
i}\right)^{1-d}\right. \nonumber \\ &+& \left.\frac{1}{2}
\Gamma(d)\left(\frac{\ln(-z)}{2\pi i}
\right)^{-d}+O\left(|\ln(z)|^{-1-d}\right)\right]\;. \label{n19}
\end{eqnarray}
Since
\begin{eqnarray}
{\cal I}_1(d) &=& M_W^2 \Gamma(1-d) z\left(1+z\frac{d}{dz}\right)
\Phi(-z,1-d,1) \nonumber \\ &=& -M_W^2 \Gamma(1-d)F(-z,-d)
\label{n20}
\end{eqnarray}
and
\begin{equation}
-F(-z,-d)=\frac{1}{\pi}e^{-i\pi d /2}\Gamma(d)(-i\ln
z)^{-d}\sin(\pi d)\;, \label{n21}
\end{equation}
Eq.~(\ref{n14}) follows at once.

\section{Appendix: Derivation of Eq.~(\ref{n23b})}

According to the modified DGLAP equations~\cite{ZAL}, we have  
\bea
\frac{df_q^{full}(x,Q^2)}{d\ln Q^2} &=& \frac{df_q(x,Q^2)}{d\ln Q^2} 
+ \frac{\alpha_s^2}{Q^2} K \left[\int^x_{x/2} \frac{dy}{y}
F_{qg}\left(\frac{x}{y}\right) \left({f_g^{full}(y,Q^2)}\right)^2
\right. \nonumber \\
&-& \left. \int_x^{1/2} \frac{dy}{y}
F_{qg}\left(\frac{x}{y}\right) \left({f_g^{full}(y,Q^2)}\right)^2
\right] \nonumber \\
&\sim & \frac{df_q(x,Q^2)}{d\ln Q^2} 
+ \frac{\alpha_s^2}{Q^2} K \left[\int^x_{x/2} \frac{dy}{y}
F_{qg}\left(\frac{x}{y}\right) f_g^2(y,Q^2)
\right.  \\ 
&-& \left. \int_x^{1/2} \frac{dy}{y}
F_{qg}\left(\frac{x}{y}\right) f_g^2(y,Q^2)
\right] \nonumber \\
&\equiv & \frac{df_q(x,Q^2)}{d\ln Q^2} 
+ \frac{\alpha_s^2}{Q^2} K R_q(x,Q^2)\;. \nonumber 
\eea
Then, using Eqs.~(28), (29) and (32) of Ref.~\cite{FAL2} and recalling
that we put $K_1=K_2$, we easily get 
\bea
R_q(x,Q^2) & \stackrel{M}{\to} & 
\Bigl[ \tilde F_{qg}^{(r)}(n) - F_{qg}^{(r)}(n) \Bigr] 
\, f_2(n) - F_{qg}^{(r)}(n) \, f_2(n) \nonumber \\
&\stackrel{n\to 1} =& \left(\frac{131}{180}-2\,\frac{1813}{2880}\right)f_2(n)
=-\frac{17}{32}f_2(n) \\
&\stackrel{M^{-1}}{\to}& -\frac{17}{32}f_g^2(x,Q^2)\;.
\eea

\vfill \eject

\end{document}